\documentclass{article}
\usepackage{graphicx}  
\usepackage{amsmath}   
\usepackage{amssymb}   
\usepackage{bm} 
\usepackage{dcolumn}
\usepackage{color}
\usepackage{mathrsfs}
\usepackage{amsfonts}
\usepackage{varioref}
\RequirePackage[colorlinks,citecolor=blue,urlcolor=magenta,linkcolor=blue]{hyperref}
\labelformat{section}{Section #1} 
\labelformat{subsection}{Section #1} 
\labelformat{subsubsection}{Section #1}
\labelformat{subsubsubsection}{Section #1}
\labelformat{equation}{Eq.~(#1)} 
\labelformat{figure}{Fig.~#1} 
\labelformat{subfigure}{Fig.~\thefigure#1} 
\labelformat{table}{Tab.~#1} 
\labelformat{appendix}{Appendix #1}

\title{\bf Light bending, static dark energy and related uniqueness of Schwarzschild-de Sitter spacetime}
\author{\bf Md Sabir Ali$^{1}$\footnote{sabir@ctp-jamia.res.in} ~and~ Sourav Bhattacharya$^{2}$\footnote{sbhatta@iitrpr.ac.in}\\
$^{1}$\small{Centre
for Theoretical Physics, Jamia Millia Islamia, New Delhi 110 025, India}\\
$^{2}$\small{Department of Physics, Indian Institute of Technology Ropar, Rupnagar, Punjab 140 001, India} }

\begin{document}
  
\maketitle

\begin{abstract}
\noindent  
Since the Schwarzschild-de Sitter spacetime is static inside the cosmological event horizon, if the dark energy state parameter is sufficiently close to $-1$, apparently one could still  expect  an effectively static  geometry, in the attraction dominated region inside the maximum turn around radius, $R_{\rm TA, max}$, of a cosmic structure.  We take the first order metric derived recently assuming a static and ideal dark energy fluid with equation of state $P(r)=\alpha\rho(r)$ as a source in Ref.~\cite{Bhattacharya:2017yix}, which reproduced the expression for $R_{\rm TA, max}$ found earlier in the cosmological McVittie spacetime. Here we show that the equality originates from the equivalence of geodesic motion in these two backgrounds, in the non-relativistic regime.  We extend this metric up to the third order and compute the bending of light using the Rindler-Ishak method. For $ \alpha\neq -1$, a dark energy dependent term appears  in the bending equation, unlike the case of the cosmological constant, $\alpha=-1$. Due to this new term in particular,  existing data for the light bending at galactic scales  yields, $(1+\alpha)\lesssim {\cal O}(10^{-14})$, thereby practically ruling out any such static and inhomogeneous dark energy fluid we started with. Implication of this result pertaining the uniqueness of the Schwarzschild-de Sitter spacetime in such inhomogeneous dark energy background is discussed.
\end{abstract}

  
\bigskip
\section{Introduction}\label{sec.1}
\noindent
One of the most remarkable  predictions of the General Theory of Relativity is the bending of a light trajectory, in stark contrast with the flat spacetime~\cite{Wald:1984rg, Paddy}. As long as the spacetime curvature is small compared to the variation of the phase of the Maxwell field, it can be shown that the phase velocity 4-vector is  a null geodesic. Thus, it is reasonable to treat the trajectory of the Maxwell field or light to  be a null geodesic at least for non-singular spacetimes, known as the null geodesic approximation~\cite{Wald:1984rg}. For the Schwarzschild spacetime in the weak gravity regime, the azimuthal deflection angle due to the bending of a light ray, whose initial and final observation points are located in the asymptotic flat region turns out  in the leading order to be $\Delta =4\pi GM/b$. Here $M$ is the mass of the gravitating object and $b=L/E$ is the impact parameter or the distance of the closest approach to  $M$, where $E$ and $L$ are respectively the conserved energy and the orbital angular momentum. For the solar system, a light trajectory which grazes the surface of the Sun, one has $\Delta  \simeq  1.75\,{\rm arc~seconds}$. A successful  detection of $\Delta$ was one of the first classic observational confirmations of the General Relativity (see e.g.~\cite{Wald:1984rg} and references therein). The aforementioned formula for light bending also passes with flying colours at the galactic length scales, see~\cite{Will} and references therein. We also refer our reader to e.g.~\cite{Bjerrum-Bohr:2014zsa, Rodrigues:2015rya} for a quantum field theoretic derivation of the same effect, by considering the gravitational analogue of the Coulomb scattering diagram of quantum electrodynamics and also for loop effects.

The observational confirmation of the accelerated expansion of our universe strongly indicates that it is endowed with the so called dark energy, see 
e.g.~\cite{Weinberg:2008zzc} and references therein. The simplest and a very successful model of the dark energy so far has been the positive cosmological constant $\Lambda$, for which a static and spherically symmetric  metric, namely the Schwarzschild-de Sitter spacetime can be derived,
\begin{eqnarray}
ds^2=-\left(1-\frac{2MG}{r}-\frac{\Lambda r^2}{3} \right)dt^2+\left(1-\frac{2MG}{r}-\frac{\Lambda r^2}{3} \right)^{-1}dr^2+r^2\left(d\theta^2+\sin^2\theta d\phi^2\right)
\label{ta0}
\end{eqnarray}
breaking the asymptotic flat boundary condition of the Schwarzschild geometry ($\Lambda=0$). The derivation of the light bending in the Schwarzschild spacetime explicitly uses the asymptotic flat boundary condition~\cite{Wald:1984rg}. Thus it becomes interesting on both physical and technical grounds, to develop a formalism to include asymptotic non-flat boundary conditions as well, in order to incorporate the effect due to $\Lambda$ or any other dark energy model, into this phenomenon. Certainly, owing to the observed tiny value of the dark energy density ($\Lambda \sim 10^{-52}\,{\rm m^{-2}}$), one should expect negligible effect due to $\Lambda$ into the light bending, if there is any,  at small scales like the solar system. However, since the $\Lambda$-term in the above metric increases monotonically with the radial distance, it is reasonable to expect that such effect could be considerable at large length scales pertaining galaxies or the clusters of galaxies.     

To the best of our knowledge, the first of the successful computations regarding the effect of the dark energy in the light bending phenomenon appeared in~\cite{Finelli:2006iz, Rindler:2007zz, Ishak:2007ea}. It was shown in ~\cite{Rindler:2007zz} that even though $\Lambda$ does not appear explicitly in the bending equation,  in the expression for the bending angle it  appears indeed, via an inner product to define the angle and the first order $\Lambda$-contribution was shown to be $-\Lambda R_0^3/6MG$, where $R_0$ is the distance of the closest approach. The minus sign signifies the repulsive effect. (We shall elucidate this so called Rindler-Ishak method, technically in \ref{sec.4}. Further in~\cite{Ishak:2007ea},
the existing lensing data for large scale structures was used in this formalism and a new, independent upper bound of the cosmological constant, $\Lambda \lesssim 10^{-50}\,{\rm m^{-2}}$, was found.

Since  this initiation, considerable efforts are being paid to pursue various aspects of this novel phenomenon, e.g.~\cite{Sereno:2007rm}-\cite{Bhattacharya:2008fu} (also references therein), with the goal to check whether we may use light bending as a direct probe to the dark energy. In particular, various aspects of gravitational weak and strong lensing in the presence of  a positive $\Lambda$ can be seen in~\cite{Sereno:2007rm}-\cite{Lebedev:2016kun} and references therein. We refer our reader to
e.g.~\cite{Kraniotis:2010gx}-\cite{Zhao:2016ltm} for discussions on light bending in de Sitter black hole spacetimes with rotation and charge. In~\cite{Villanueva:2013gga, Olivares:2016zqq, Lim:2016lqv}, discussions of this phenomenon in the context of conformal and some other modified gravity theories can be found. See~\cite{Virbhadra:2007kw} for a discussion on the effect of a naked curvature singularity on light bending. We also refer our reader to~\cite{Bhattacharya:2008fu} for a derivation of the bending angle in the de Sitter cosmic string background. 

On the other hand, however, there has been considerable criticism to Rindler-Ishak's formalism, e.g.~\cite{Khriplovich:2008ij, Simpson:2008jf, Butcher:2016yrs, Piattella:2015xga}, as well. The chief  objection raised is that, if one uses the cosmological McVittie description instead of the static geometry, there will be no light bending due to $\Lambda$ at all, in the leading order.   However, recently in~\cite{Faraoni:2016wae}, a non-vanishing contribution of $\Lambda$ has been claimed for the McVittie spacetime, by using an effective local mass function. Thus it would be fair  to admit that the debate is still far from being over.  \\

 \noindent
Can we use this phenomenon to probe or constrain possible inhomogeneity of the dark energy in a static background, pertaining isolated structures decoupled from universe's expansion? 
Different  static descriptions of spacetimes with nontrivial dark energy in the context of light bending was earlier attempted   in e.g.~\cite{Finelli:2006iz, Fernando:2014rsa, He:2017alg} also. See also~\cite{Chamseddine:2016uyr, Kluson:2017iem} (also~\cite{Sebastiani:2016ras} for a recent review and list of references) for a discussion on the so called mimetic model of the inhomogeneous dark energy. 

 Even though for the dark energy equation of state $P(t)=w\rho(t)$ with $w\neq -1$, there is no notion of a static spacetime geometry as such, we might  expect {\it a priori } such a description to hold effectively  up to a certain length scale,  as follows. Since  the positive $\Lambda$ fits various data exceedingly well~\cite{Weinberg:2008zzc}, we may  imagine  $w$ to be sufficiently close to minus of unity. Since with a positive $\Lambda$  we indeed have a static description of our spacetime, \ref{ta0}, up to the cosmological event horizon, it seems  reasonable to expect that it will be so for $w$ sufficiently close to minus of unity as well effectively, but up to  some relevant length scale {\it much smaller} than the horizon. That length scale naturally seems to be the maximum turn around radius $R_{\rm TA,max}$ (also sometimes called the static radius) --  the point where the attraction due to the mass  $M$ of a spherical structure gets precisely balanced with the repulsion due to the ambient dark energy~\cite{Stuclik1, Stuclik2, Pavlidou:2013zha, Pavlidou:2014aia}. Note that this is analogous to  the vacuole model of a structure inside the Einstein radius discussed in e.g.~\cite{Ishak:2007ea, Bhattacharya:2010xh} and references therein, for $\Lambda{\rm CDM}$.
 
In the region $r\leq R_{\rm TA, max}$, we shall consider an effective  static and spherically symmetric description of the spacetime with a dark energy fluid with equation of state $P_E(r)= \alpha\rho_E(r)$.  There is no reason to imagine  that such description of the dark energy is a fundamental one. In other words, with such effective description, we {\it cannot}
possibly assert that $\alpha$ equals the state parameter $w$ of the homogeneous description of the dark energy, $P_E(t)=w\rho_E(t)$, holding beyond the maximum turn around radius.

 It was shown in~\cite{Pavlidou:2013zha, Pavlidou:2014aia} that for $\Lambda{\rm CDM}$, the computation of $R_{\rm TA, max}$ yields the same result via both time independent Schwarzschild-de Sitter spacetime, \ref{ta0}, and the time dependent cosmological scalar perturbation theory,
$$R_{\rm TA, max}=\left(\frac{3MG}{\Lambda}\right)^{1/3}$$
In~\cite{Pavlidou:2014aia}, this result was also extended to the dark energy state parameter  $w\neq -1$, using cosmological scalar perturbation theory, yielding
\begin{eqnarray}
R_{\rm TA, max}= \left(\frac{-6MG}{\Lambda(1+3w)}\right)^{1/3} \qquad ({\rm with }~~ w<-1/3)
\label{ta0'}
\end{eqnarray}
where setting $w=-1$ recovers the $\Lambda{\rm CDM}$ result. Can we derive this result, using a static geometry as well, just like in $\Lambda{\rm CDM}$? Indeed very recently in~\cite{Bhattacharya:2017yix}, a static and spherically symmetric  linearized metric using the aforementioned  effective static and  dark energy fluid as the source was derived, 
\begin{eqnarray}
ds^2\approx-\left(1-\frac{2M G}{r}+ \frac{(1+3\alpha)\Lambda r^2}{6} \right)dt^2+ \left(1+\frac{2M G}{r}+\frac{\Lambda r^2}{3}\right)dr^2 +r^2d\Omega^2
\label{ta0''}
\end{eqnarray}
Setting $\alpha=-1$ recovers \ref{ta0} in the leading order, in which case the constant $\Lambda$ is interpreted as the cosmological constant. The maximum turn around radius for this metric corresponds to the maximum of the norm, $f\, (\equiv -g_{tt})$, of the timelike Killing vector field ($\partial_r f=0$)~\cite{Pavlidou:2013zha}, giving $R_{\rm TA, max}= (-6MG/\Lambda(1+3\alpha))^{1/3}$, formally similar to \ref{ta0'}.  We shall see that this equality basically originates  from the equivalence the equation of motion of a non-relativistic test particle in \ref{ta0''} and in the cosmological  McVittie spacetime, \ref{sec.2}. These two results certainly motivate us to take a further look into \ref{ta0''}. \\

\noindent
Since setting $\alpha=-1$ in \ref{ta0''} recovers \ref{ta0},  the staticity of the dark energy for $\alpha \neq -1$ should be analogous to possible inhomogeneity (i.e. $\rho\equiv \rho(r)$) in it.  The crucial task now is to precisely  find a bound on such inhomogeneity and to distinguish between the static and time dependent, spatially homogeneous  descriptions of the dark energy. We note that a possible caveat in  any inhomogeneous dark energy fluid model could be the local instability arising from the imaginary sound speed. Let us imagine a  perturbation is built in a small spacetime region. Due to the imaginary sound speed, the perturbation cannot propagate and it might eventually create an instability by growing with time. However, as long as $\alpha$ is sufficiently `close' to $-1$, it seems fair to expect such effects to be not significant, in a reasonable timescale. Nevertheless, we may avoid such instability at sufficiently late times as well,  by modifying the equation of state as follows. The growth of the perturbation with time would also backreact and the fluid would eventually become time dependent. We may modify the effective fluid description now to be 
$$P_E(r,t)=\left( \alpha+ \delta \alpha(r,t)\right) \rho_E(r,t)$$
So that we have the adiabatic sound speed, 
$$c_a^2:=\frac{\dot P_E}{\dot \rho_E}=\alpha+ \rho_E\frac{\delta{\dot \alpha}}{\delta {\dot \rho_E}} $$
Note that this is analogous to what is done in order to define a sound speed  in  homogeneous cosmological models as well, 
e.g.~\cite{Hannestad}, by allowing the state parameter to vary with time. Now when the perturbation grows big, we can expect the second quantity on the right hand side to be comparable with the first term which is negative. If the second quantity is positive, we might expect a non-vanishing sound speed,
through which the dark energy perturbation would propagate, eventually  taming down the local instability. Nevertheless, it would turn out that such inhomogeneity in the dark energy we start with is practically ruled out, as follows.

We shall further extend \ref{ta0''} up to the third order in \ref{sec.3}. Note that $g_{tt}\cdot g_{rr}\neq -1$ for \ref{ta0''}, consistent with the general theorem established in~\cite{Jacobson:2007tj}, that
for the equality one must have $T_t{}^t+T_r{}^r=0$ (\ref{sec.3}). In \ref{sec.4}, we shall derive the bending equation for light and will see that unlike the $\alpha=-1$ case~\cite{Rindler:2007zz}, a term containing the dark energy  shows  up in the bending equation (\ref{ta13'}), originating from that inequality. We shall further show in \ref{sec.4.1} that due to this  term in particular, the existing  data for the light bending at galactic  scales leads to a severe constraint : $(1+\alpha) \lesssim 10^{-14}$. This precise number is the main result of this paper and such astonishing smallness was certainly not obvious {\it a priori}. It shows that, {\it inside} the maximum turn around radius of a structure, as far as obtaining a static spacetime geometry is concerned in an effective static and ideal dark energy fluid background,   the Schwarzschild-de Sitter, \ref{ta0}, is the only acceptable solution. We finally conclude in \ref{sec.5}.

We shall use mostly positive signature for the metric ($-,+,+,+$) and will set $c=1$ throughout. We shall {\it always} implicitly assume below that the dark energy is `close' to a positive $\Lambda$, so that our above assertion of static description inside $R_{\rm TA, max}$ expectedly remains justified. 

\section{Equivalence of non-relativistic motion in \ref{ta0''} and McVittie spacetime}\label{sec.2}
\noindent
The maximum turn around region is a non-relativistic regime for a massive test particle where its acceleration vanishes~\cite{Pavlidou:2013zha}. We shall show below that the equality of $R_{\rm TA, max}$ found using \ref{ta0''}
and using cosmological framework basically originates from the equivalence of the geodesic equation in the non-relativistic regime
in \ref{ta0''} and in the cosmological McVittie spacetime.

We start with the geodesic equation, 
$$\frac{d^2x^{\mu}}{d\lambda^2}+\Gamma^{\mu}_{\nu \rho} \frac{dx^{\nu}}{d\lambda}\frac{dx^{\rho}}{d\lambda}=0,$$
and consider  \ref{ta0''} first. Due to the spherical symmetry, we can set $\theta =\pi/2$. Then in the week field limit we are interested in, the leading order radial equation becomes
\begin{eqnarray}
\frac{d^2r}{dt^2}+\frac{GM}{r^2}+\frac{(1+3\alpha)\Lambda r}{6}-\frac{L^2}{r^3}=0
\label{nr1}
\end{eqnarray}
where we have set the proper time to be equal to the coordinate time in the non-relativistic limit, $\lambda \sim t$, and the radial velocity $dr/d\lambda$ in this limit is much small compared to unity and also, $\Gamma^r_{tt}\approx GM/r^2+(1+3\alpha)\Lambda r/6$. We also have used the conserved orbital angular momentum, $L= r^2 d\phi/dt$. The maximum turn around condition is obtained by setting $d^2r/dt^2=0$ for radial geodesics ($L=0$).

The  McVittie metric, on the other hand,  reads at the linear order, in the absence of any off-diagonal spatial stresses~\cite{Weinberg:2008zzc}, 
\begin{eqnarray}
ds^2= -\left(1-2\Psi(R,t)\right)dt^2+\left(1+2\Psi(R,a(t))\right)a^2(t)\left[dR^2+R^2d\Omega^2\right]
\label{nr2}
\end{eqnarray}
where we have the Newton potential, $\Psi=G{\widetilde M}/Ra(t)$ and $a(t)$ is the scale factor. The dark energy is specified as
$$T^E_{00}=\rho_E(t) \qquad T^E_{ij}= a^2(t)P_E(t)\delta_{ij}$$
whereas the cold dark matter energy momentum tensor is given by,
$$T^M_{00}=\rho_m(t)\qquad T^M_{ij}=0$$

We take the equation of state $P_E(t)=w\rho_E(t)$, to have for the background homogeneous spacetime,
\begin{eqnarray}
\rho_E(t)=\rho_{0E}a^{-3(1+w)}(t) \qquad \rho_m(t)= \rho_{0m}a^{-3}(t)
\label{nr3}
\end{eqnarray}
where $\rho_{0E}$ and $\rho_{0m}$ are integration constants. We have from the spatially homogeneous Einstein's equations
\begin{eqnarray}
H^2(t)=\frac{\dot{a}^2}{a^2}=\frac{8\pi G}{3}\left(\frac{ \rho_{0E}}{a^{3(1+w)}(t)}+ \frac{\rho_{0m}}{a^{3}(t)}\right),
\quad{\dot H}(t)=-4\pi G\left(\frac{ \rho_{0E}(1+w)}{ a^{3(1+w)}(t)} +\frac{ \rho_{0m}}{ a^{3}(t)}\right)
\label{nr4}
\end{eqnarray}
Defining now the proper radial coordinate, $r=a(t)(1+\Psi)R$, we rewrite \ref{nr2} as 
\begin{eqnarray}
ds^2\approx-\left(1-2\Psi-H^2(t)r^2\right)dt^2 +\left(1+2\Psi\right)dr^2-2H(t)rdrdt+r^2d\Omega^2
\label{nr5}
\end{eqnarray}
where since we are interested in the subhorizon length scale, we have ignored the temporal variation of the potential $\Psi$ with respect to its spatial variation. The equation of motion for a non-relativistic particle in the above background becomes
\begin{eqnarray}
\frac{d^2 r}{dt^2}+ \left(\frac12 g^{tr}\partial_t g_{tt}+g^{rr}\partial_t g_{tr}-\frac12 g^{rr}\partial_r g_{tt}\right)-\frac{L^2}{r^3}=0
\label{nr6}
\end{eqnarray}
which becomes in the leading order, after computing the connections from \ref{nr5}
\begin{eqnarray}
\frac{d^2 r}{dt^2}- r({\dot H} +H^2 ) -\partial_r \Psi-\frac{L^2}{r^3}=0
\label{nr7}
\end{eqnarray}
which, after using \ref{nr4} and the expression for $\Psi$ becomes
\begin{eqnarray}
\frac{d^2 r}{dt^2}+ \frac{ 8\pi G \rho_{0E} \,(1+3w) r}{6 a^{3(1+w)}(t)} + \frac{G(\widetilde{M}+\delta {\widetilde M}(t))}{r^2}-\frac{L^2}{r^3}=0
\label{nr8}
\end{eqnarray}
where 
$$\delta {\widetilde M}(t): = \frac{4 \pi \rho_m(t) r^3}{3} $$
is the contribution of the homogeneous cold dark matter to the mass.

Thus the formal equivalence of \ref{nr1} and \ref{nr8} is apparent via the following  identifications of the observable quantities at a given cosmological time $t=t_0$ : $$M\equiv \widetilde{M}+\delta {\widetilde M}(t=t_0) \qquad {\rm and} \qquad\Lambda \equiv  \frac{8\pi G \rho_{0E}}{a^{3(1+w)}(t=t_0)}$$
just because in \ref{ta0}, \ref{ta0''} or \ref{nr1}, we are using the {\it current  observed} values of the dark energy density  and the mass and the time scale of observing a test particle is certainly much smaller than the time required to change the scale factor and hence the  dark energy/dark matter density to any considerable extent.{\footnote{Note that since the scale factor is dimensionless, it must be  a function of $\sim \sqrt{\Lambda} t$. Thus for any interval of observation of the cosmic time from  $t=t_0$ to $t=t_0+T$ with $T\ll t_0$ as $t_0$ is the age of the universe, the scale factor practically remains unchanged.}}

Thus both \ref{ta0''} and \ref{nr5} predict equivalent leading order non-relativistic geodesic trajectory, such as a star orbiting the galactic centre. This serves as the key motivation to the computations  we make below.

\section{The metric  in higher perturbative orders}\label{sec.3}
\noindent
Let us start with the ansatz for a static and spherically symmetric spacetime,
\begin{eqnarray}
ds^2=-f(r)dt^2+h(r)dr^2+r^2\left(d\theta^2+\sin^2\theta d\phi^2\right)
\label{ta1}
\end{eqnarray}
In this background, we may take the energy momentum tensor for the dark energy fluid as,
\begin{eqnarray}
T^E_{ab}=\rho_{E}(r)\, u_au_b+P_E(r)\,\left(g_{ab}+u_a u_b\right),
\label{ta2}
\end{eqnarray}
where $\rho_E$, $P_E$ and $u_a$ are respectively the energy density, pressure and the velocity of the fluid's world line. For a static geometry,  the  world lines $u^a$ of a backreacting fluid must be parallel to the orbits of the timelike Killing vector field. Thus, normalizing $u_au^a=-1$,  we can take $u^a=f^{-1/2}(\partial_t)^a$. We assume the equation of state $P_E= \alpha\rho_E$ with $\rho_E>0$ where $\alpha$ is a constant. Thus we have $$R_{ab}u^au^b =8\pi G \left(T^E_{ab}-\frac12 T^E g_{ab}\right)u^au^b= (1+3\alpha)8\pi G\rho_E(r)$$ so that in order to violate the strong energy condition  or to generate the repulsive effect~\cite{Wald:1984rg}, we must have $\alpha<-1/3$. The Einstein equations for this system are given  by
\begin{eqnarray}
\frac{h'}{h^2 r}+\frac{1}{r^2}-\frac{1}{hr^2}=8\pi G \rho_E(r),  \qquad
\frac{f'}{hfr}+\frac{1}{hr^2}-\frac{1}{r^2}=8\pi G \alpha \rho_E(r)\nonumber\\
\frac{f''}{2fh}-\frac{h'}{2h^2 r}+\frac{f'}{2fhr}-\frac{f'h'}{4h^2f}-\frac{f'^2}{4f^2 h}=8\pi G \alpha \rho_E(r)
\label{ta3}
\end{eqnarray}
where a `prime' denotes differentiation once with respect to the radial coordinate. It is easy to find, expanding the conservation equation, $\nabla^a T^{E}_{ab}=0$, 
\begin{eqnarray}
\frac{(1+\alpha)}{2\alpha f}\frac{df}{dr}+\frac{1}{\rho_E}\frac{d\rho_E}{dr}=0,
\label{ta3''}
\end{eqnarray}
which yields
\begin{eqnarray}
\rho_E(r)=\frac{\rho_{0}}{f^{\frac{1+\alpha}{2\alpha}}}
\label{ta3'}
\end{eqnarray}
where $\rho_{0}$ is an integration constant. The above relation is an analogue 
of the time dependence of the energy density, $\rho_E\sim a^{-3(1+w)}(t)$ in a spatially homogeneous cosmological spacetime, where the scale factor $a(t)$ is equivalent to the redshift or the Tolman factor 
$f(r)$, for a static spacetime.  

With the help of \ref{ta3''} and \ref{ta3'}, \ref{ta3} can be solved perturbatively as follows, starting from the maximum turn around radius, $R_{\rm TA,max}$. This point corresponds to the maximum of the norm of the timelike Killing vector field, $f'(r)=0$, where the acceleration corresponding to the orbits of the timelike Killing vector field,
$$(\partial_t)^a\nabla_a (\partial_t)^b= \frac12 g^{ab}\nabla_b f \equiv \frac{f'}{2h}$$
vanishes. \ref{ta3''} then shows that we may ignore the variation of $\rho(r)$ at or around the maximum turn around radius.  Then writing $\rho_E=\rho_0~({\rm const.})$ and setting $8\pi G\rho_0=\Lambda$ into the Einstein equations, we arrive at the first order metric in \ref{ta0''} ~\cite{Bhattacharya:2017yix}. Also, by  adding the first two of \ref{ta3}, we get
$$ \partial_r \ln (fh) =\frac{\Lambda (1+\alpha) rh }{f^{\frac{1+\alpha}{2\alpha}}}$$
showing that $g_{tt}\cdot g_{rr}=-1$ is possible only with either $\alpha=-1$ or $\Lambda=0$. In other words, since  $T^E_t{}^t+T^E_r{}^r= (1+\alpha ) \rho_E (r)\neq 0$ for $\alpha\neq -1$, we have  $g_{tt}\cdot g_{rr}= -(1+ (1+\alpha)\Lambda r^2/2)$ for \ref{ta0''}, in accordance with the general theorem of~\cite{Jacobson:2007tj}.

Now, for structures with mass starting from a few solar mass up to as large as $10^{17}M_{\odot}$, it is easy to see that the ${\cal O}(GM/r)$ and ${\cal O} (\Lambda r^2) $ terms are much smaller compared to unity in the maximum turn around region.  Thus setting $\rho$ or $P$
as constants in \ref{ta3} could also be interpreted just as -- keeping only the leading order terms in the source.

We can continue the perturbative iteration now, in order to extend the above metric off the maximum turn around length scale. In the next order of the perturbation theory, clearly, we have to take $f(r)$ from \ref{ta0''} and substitute 
$$\rho_E(r)=\frac{\rho_{0}}{f^{\frac{1+\alpha}{2\alpha}}(r)} \approx \rho_0 \left(1+\frac{GM(1+\alpha)}{\alpha r}-\frac{(1+\alpha)(1+3\alpha)\Lambda r^2}{12 \alpha}\right) $$   
on the right hand side of \ref{ta3}, to obtain the second order perturbative metric 
\begin{eqnarray}
&&f(r)\approx \left(1-\frac{2M G}{r}+ \frac{(1+3\alpha)\Lambda r^2}{6} + \frac{GM \Lambda r (4 \alpha^2 +5\alpha+1)}{2\alpha}+ \frac{\Lambda^2 r^4(3\alpha+51\alpha^2+45\alpha^3-3)}{720 \alpha}\right)\nonumber\\
&&h(r)\approx \left(1+\frac{2M G}{r}+\frac{\Lambda r^2}{3} +\frac{4G^2 M^2}{r^2}+ \frac{GM \Lambda r (11 \alpha+3)}{6\alpha} +\frac{\Lambda^2 r^4 (8\alpha-3-9\alpha^2)}{180\alpha}\right)
\label{ta8}
\end{eqnarray}
Setting $\alpha=-1$ above again recovers the Schwarzschild-de Sitter spacetime \ref{ta0}, expanded up to the second order. Likewise for the third order, using \ref{ta8} we get
$$\rho_E(r)\approx \rho_0\left(1+\frac{GM(1+\alpha)}{\alpha r}- \frac{(1+4\alpha+3\alpha^2) \Lambda r^2}{12 \alpha}+ \frac{G^2M^2 (1+4 \alpha+3\alpha^2)}{2\alpha^2 r^2}\right.$$ $$\left.-\frac{GM\Lambda r(1+\alpha)(21\alpha^2+21\alpha+4)}{12\alpha^2}
+\frac{\Lambda^2 r^4 (1+\alpha) (4+21\alpha+42\alpha^2+45\alpha^3)}{720 \alpha^2}\right)$$
and substitute it into the Einstein equations, to obtain
\begin{eqnarray}
&&f(r)\approx \Bigg(1-\frac{2M G}{r}+ \frac{(1+3\alpha)\Lambda r^2}{6} + \frac{GM \Lambda r (4\alpha^2 +5\alpha+1)}{2\alpha}+ \frac{\Lambda^2 r^4(3\alpha+51\alpha^2+45\alpha^3-3)}{720 \alpha}\nonumber\\
&&+\frac{G M\Lambda^2 r^3(90\alpha^4+129\alpha^3+28\alpha^2-15\alpha-4)}{144\alpha^2}+\frac{\Lambda^3 r^6(27\alpha^4+6\alpha^3-28\alpha^2-6\alpha+1)}{7560\alpha^2}\Bigg)\nonumber\\
&&h(r)\approx \Bigg(1+\frac{2M G}{r}+\frac{\Lambda r^2}{3} +\frac{4G^2 M^2}{r^2}+ \frac{GM \Lambda r (11\alpha+3)}{6\alpha} +\frac{\Lambda^2 r^4 (8\alpha-3-9\alpha^2)}{180\alpha}\nonumber\\
&&+\frac{8G^3M^3}{r^3}+\frac{G^2M^2\Lambda(15\alpha^2+8\alpha+1)}{2\alpha^2}-\frac{GM\Lambda^2r^3(153\alpha^3+34\alpha^2+61\alpha+20)}{240\alpha^2}\nonumber\\
&&+\frac{\Lambda^3 r^6(135\alpha^4-243\alpha^3+77\alpha^2-93\alpha+12)}{15120\alpha^2}\Bigg)
\label{add1}
\end{eqnarray}
Setting $\alpha=-1$ above once again recovers the Schwarzschild-de Sitter spacetime, expanded up to the third order. In principle one can continue such iteration up to arbitrary order of the perturbation but for our current purpose the above would be suffice. Note that since the above metric was derived starting from $R_{\rm TA,max}$, the higher order terms in the mass, in particular, becomes relevant as we move inward. On the other hand,  such static scenario certainly cannot be appropriate for length scales  larger than $R_{\rm TA,max}$, for such length scale falls essentially into the realm of the accelerated expansion.   

Moreover, not all the terms appearing in \ref{add1} would be phenomenologically relevant, keeping in mind that we are working at length scales much smaller than the Hubble radius ($\sim 1/\sqrt{\Lambda}$). We thus would retain and work with only the following phenomenologically leading terms in the metric function,   
\begin{eqnarray}
&&f(r)\vert_{\rm effective}\approx \left(1-\frac{2M G}{r}+ \frac{(1+3\alpha)\Lambda r^2}{6} + \frac{GM \Lambda r (4\alpha^2 +5\alpha+1)}{2\alpha}\right)\nonumber\\
&&h(r)\vert_{\rm effective}\approx \left(1+\frac{2M G}{r}+\frac{4G^2 M^2}{r^2}+\frac{8G^3M^3}{r^3}+\frac{\Lambda r^2}{3} + \frac{GM \Lambda r (11\alpha+3)}{6\alpha} \right)
\label{add2}
\end{eqnarray}
The ${\cal O}(\Lambda r^2)$ terms contribute to the leading order of the light bending due to the dark energy, as discussed in~\cite{Rindler:2007zz, Ishak:2007ea} and in many other places for $\alpha=-1$. The ${\cal O}(GM\Lambda r) $ terms, being much less than $\Lambda r^2 $, should be regarded as the next to leading order effect into this phenomenon, both due to the dark energy and the mass. We shall use below these  effective metric functions only, to compute the effect of the dark energy in the bending of light.

\section{Light bending  with $\alpha\neq -1$ and consistency with solar system data}\label{sec.4}
Let us first construct the equation for a geodesic $u^a\equiv dx^a/d\lambda$, where $\lambda$ is an affine parameter along the geodesic, travelling in the background of \ref{ta1}. We take the norm  $u^au_a=-\kappa$ where $\kappa = 0\,(1)$
for null (timelike) trajectory. Owing to the  spherical symmetry of the spacetime, we set $\theta=\pi/2$  and obtain 
\begin{eqnarray}
-f(r)\left(\frac{dt}{d\lambda}\right)^2+h(r)\left(\frac{dr}{d\lambda}\right)^2 + r^2 \left(\frac{d\phi}{d\lambda}\right)^2=-\kappa.
\label{ta9}
\end{eqnarray}
The time translation Killing vector field $(\partial_t)^a$ and the rotational Killing vector field $(\partial_{\phi})^a$ admits two conserved quantities -- the energy $(E)$ and the orbital angular momentum $(L)$~\cite{Wald:1984rg}, 
\begin{eqnarray}
E=-g_{ab}(\partial_t)^au^b= f(r)\frac{dt}{d\lambda} \qquad L= g_{ab} (\partial_{\phi})^a u^b= r^2 \frac{d\phi}{d\lambda}
\label{ta10}
\end{eqnarray}
Using \ref{ta10} into \ref{ta9}, we get
\begin{eqnarray}
\frac{dr}{d\lambda} =\frac{1}{\sqrt{fh}} \left[E^2-f\left(\frac{L^2}{r^2}+\kappa\right)\right]^{1/2} 
\label{ta11}
\end{eqnarray}
The qualitative difference of geodesic  motion  between spacetimes with $f\cdot h \neq 1$   with that of $f\cdot h =1$ is apparent here -- for the former case, we have an energy dependent effective potential term, like that of the stationary axisymmetric spacetimes, see e.g. chapter 12 of~\cite{Wald:1984rg}.

Combining the above equation with the second of \ref{ta10} and defining the customary new variable $r=1/u$ we find for the null geodesic $(\kappa=0)$,
\begin{eqnarray}
\frac{du}{d\phi} = -\frac{\left[E^2- L^2 u^2 f(u)\right]^{1/2}}{L \sqrt{f(u) h(u)}} 
\label{ta12}
\end{eqnarray}
Differentiating once again with respect to $\phi$, we obtain 
\begin{eqnarray}
\frac{d^2u}{d\phi^2} = -\frac{\partial_u (f u^2)}{2fh} -\frac{\left(E^2/L^2-fu^2\right) \partial_u(fh)}{2  (fh)^2}
\label{ta13}
\end{eqnarray}
where $f$ and $h$ are understood as functions of $u$. Substituting for the third order metric functions from \ref{add2} and ignoring higher order terms due to $\Lambda$, we get 
\begin{eqnarray}
\frac{d^2u}{d\phi^2}+ u= 3MG u^2+ \frac{\Lambda (1+\alpha)}{2b^2u^3}+ \frac{3G M \Lambda(2\alpha^2+3\alpha+1)}{4\alpha} 
\label{ta13'}
\end{eqnarray}
where we have written $L/E=b$, the impact parameter. Setting $\alpha=-1$ recovers the equation for the Schwarzschild-de Sitter background~\cite{Rindler:2007zz}. 

We note here the chief differences between our approach and  an analogous recent attempt for light bending computation with an effective  static  dark energy~\cite{He:2017alg}. First, unlike ours \ref{ta0''} or~\ref{add2} the metric of~\cite{He:2017alg} has $g_{tt}\cdot g_{rr}=-1$. As a result
the dark energy dependent term in the bending equation \ref{ta13'} is different from ours there. Also, the maximum turn around radius for $\alpha\neq -1$ found in that reference (called there the `critical radius') is formally different from what we obtain in \ref{ta0'} or equivalently in the context of the cosmological scalar perturbation theory, in~\cite{Pavlidou:2014aia}.

Before we solve the above equation, let us first set $\alpha=-1$ and review briefly the formalism of~\cite{Rindler:2007zz},
\begin{eqnarray}
\frac{d^2u}{d\phi^2}+ u= 3MG u^2
\label{ta14}
\end{eqnarray}
The homogeneous part of he above equation has harmonic solutions which we take to be $\sin\phi/R_0$ where the constant $R_0$   is the distance of the closest approach ($\phi=\pi/2$ in fig.~1). This corresponds to setting $dr/d\lambda=0$ in \ref{ta11}, which in the leading order becomes $R_0=L/E=b $. Clearly, the homogeneous solution corresponds to the undeflected light traveling in the absence of any gravity. The choice of the $\sin$-function guarantees that $\phi\to 0,~2\pi$ as $r\to \infty$.  Even though such asymptotic limit makes sense only with $\Lambda =0$, taking solutions satisfying such limit ensures that we recover the correct Schwarzschild limit when we set $\Lambda=0$.  

In order to get some feasible effect in light bending due to the dark energy, owing to its tiny observed value, we must go away from the Schwarzschild radius of the central object, making the right hand side of \ref{ta14} `small'. Thus the idea becomes to substitute the first order homogeneous solution on the right hand side and solve for the full inhomogeneous equation and continuing the perturbative iteration procedure likewise.  

Thus, substituting $u=\sin\phi/R_0$ on the right hand side of \ref{ta14}, taking a trial solution of the form $u=A+\sin\phi/R_0+B\sin^2\phi$ where $A,~ B$ are constants   gives the next to the leading order solution
\begin{eqnarray}
u=\frac{\sin\phi}{R_0}+\frac{3GM}{2R_0^2}\left(1+\frac{\cos2\phi}{3}\right)
\label{ta15}
\end{eqnarray}
Likewise after substituting 
$$u^2 \approx \frac{\sin^2\phi}{R_0^2}  + \frac{3MG \sin\phi}{R_0^3}\left(1 +\frac{\cos2 \phi}{3}\right) $$
we get the second order solution~\cite{Ishak:2007ea}
\begin{eqnarray}
u= \frac{\sin\phi}{R_0}+\frac{3GM}{2R_0^2}\left(1+\frac{\cos2\phi}{3}\right)+\frac{3M^2G^2}{16R_0^3}\left(10\pi \cos\phi -20\phi \cos\phi -\sin3\phi\right)
\label{ta16}
\end{eqnarray}
The Rindler-Ishak method to compute the bending angle goes as follows~\cite{Ishak:2007ea}. 
\begin{figure}[h]
\begin{center}
\includegraphics[width=5.5in,height=2.0in,angle=0]{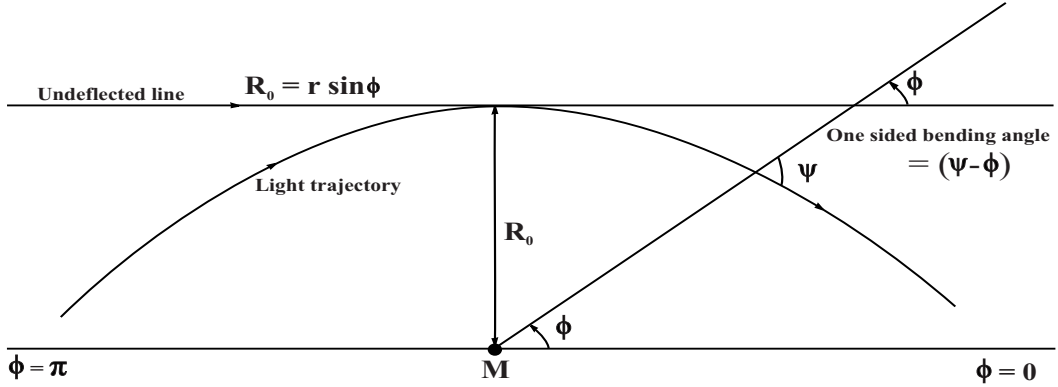}
\caption{
Schematic diagram representing the bending of light. The one-sided bending angle is $(\psi-\phi)$. Full bending angle is just twice of that, owing to the reflection symmetry with respect to the line $\phi=\pi/2$.  
}
\end{center}
\label{fig}
\end{figure}
Let $\psi$ be the angle (fig.~1) between the $\phi={\rm constant}$ line and the trajectory, given  by 
\begin{eqnarray}
\cos\psi =\frac{\gamma_{ij} n^i l^j  }{ ( \gamma_{ij} n^i n^j)^{1/2}   (\gamma_{ij} l^i l^j)^{1/2}     }
\label{ta17}
\end{eqnarray}
where $\gamma_{ij}$ is the induced metric on the $(r,~\phi)$ plane ($\equiv h(r)dr^2+r^2d\phi^2$), $n^i$ and $l^i$ are respectively the tangent vectors along the spatial part of the trajectory and the $\phi={\rm constant }$ lines. We differentiate  \ref{ta15} with respect to $\phi$  to get 
\begin{eqnarray}
\frac{dr}{d\phi} =\frac{r^2 \cos\phi}{R_0}\left(\frac{2GM}{R_0}\sin\phi-1\right)
\label{ta18}
\end{eqnarray}
We have also 
\begin{eqnarray}
n^i \equiv \{dr,~d\phi \}=\left\{\frac{dr}{d\phi},~1 \right\}d\phi \qquad l^i \equiv \{0,~1\}d\phi
\label{ta18'}
\end{eqnarray}
Then it turns out that~\cite{Rindler:2007zz} 
\begin{eqnarray}
\cos\psi = \frac{|dr/d\phi|}{\left[ (dr/d\phi)^2 +r^2/h(r) \right]^{1/2}}
\label{ta19}
\end{eqnarray}
and 
\begin{eqnarray}
\tan\psi = \frac{r}{\sqrt{h} |dr/d\phi| }
\label{ta19'}
\end{eqnarray}
For small bending angle, $\tan\psi \approx \psi$. One then puts $\phi\approx 0$ for large distance away from the source, for which from \ref{ta15} $r=R_0^2/2MG$ and finally multiples with $2$ to get the full bending angle, fig.~1,
\begin{eqnarray}
\Delta=2(\psi-\phi) \approx 2\psi= \frac{r}{\sqrt{h} |dr/d\phi| }\Bigg\vert_{\phi=0}\approx \frac{4MG}{R_0}-\frac{\Lambda R_0^3}{6MG}
\label{ta20}
\end{eqnarray}

\bigskip
\noindent
Let us come to \ref{ta13'} now. The leading term is the first one on the right hand side and the second term is the leading contribution from the dark energy, while the third term is clearly much tinier than both. Note that this particular term, being a constant, would appear in the expression for $u(\phi)$ just as an additive constant. Accordingly, this would not contribute to the light bending, \ref{ta19}, which is determined by $dr/d\phi$.   We emphasize here the qualitative difference between  the Schwarzschild-de Sitter spacetime and \ref{add2} -- we have a  direct and impact parameter dependent contribution due to the dark energy into the bending equation, \ref{ta13'}. We may compare the mass term with that of the second appearing on the right hand side of that equation in the solar system, in order to have a feel. For the solar system, a light ray that gradges the surface of the Sun, the bending angle is predicted and observationally verified  to be around $1.75$ arc seconds~\cite{Wald:1984rg}. Equating this with $4MG/b$ and using $MG \sim 10^3\, {\rm m}$ for the Sun, we obtain $b\sim 10^7\,{\rm m}$, which is, of course, approximately the radius of the Sun. Then for $\Lambda \sim 10^{-52}{\rm m}^{-2}$, we can see that the mass term, at the length scale of the closest approach, is about $10^{18}$ times larger than the dark energy term. However, this would not be the case at large length scales like galaxies or the cluster of galaxies for $\alpha\neq -1$ leading to severely constraining $\alpha$ to minus unity, which we shall prove below.

We shall solve \ref{ta13'}, perturbatively, assuming quite naturally that the dark energy's contribution should be small compared to the mass term. While the contribution due to the mass term is obviously the same as \ref{ta15} or~\ref{ta16}, we plug \ref{ta15} into the dark energy term of \ref{ta13'}. We could not find a closed form solution to this and solved it for $\phi \ll 1$, corresponding to the leading contribution of the dark energy
\begin{eqnarray}
u\vert_{\phi \ll 1}({\rm dark~energy})= \frac{3\Lambda (1+\alpha) R_0^7 }{128 b^2 (GM)^4} \left(2\phi \cos\phi-\pi \cos\phi  \right) +~ {\rm higher~order~terms} 
\label{ta21}
\end{eqnarray}
Note that the term proportional to $\pi \cos\phi$ is just the solution of the homogeneous part $d^2u/d\phi^2+u=0$. We have added this term, to make $u$ symmetric with respect to the line $\phi=\pi/2$.  Putting these all in together, and using $\sin\phi \approx \phi$, $\cos\phi \approx 1$ as $\phi \ll 1$, we have  
\begin{eqnarray}
u&=&\frac1r\approx \frac{\phi}{R_0}+\frac{2GM}{R_0^2}+\frac{3M^2G^2}{16R_0^3}\left(10\pi  -23\phi \right)+\frac{3\Lambda (1+\alpha) R_0^7 }{128 b^2 (GM)^4} \left(2\phi -\pi   \right)\nonumber\\
&+&\frac{3G M \Lambda(2\alpha^2+3\alpha+1)}{4\alpha} 
\label{ta22}
\end{eqnarray}
Differentiating this once with respect to $\phi$, substituting then into \ref{ta19'} and using \ref{add2} we can obtain $\tan \psi$. Also, as we have seen (discussions below \ref{ta14}, at the leading order $R_0=b$. In the next order, it is easy to find by setting $dr/d\lambda=0$ in \ref{ta11} that $R_0 \approx b-MG$. The actual sizes of non-black hole large scale structures are much larger than their Schwarzschild radii and the minimum value of $R_0$ must be the actual size itself. In other words, $R_0,\,b \gg MG$  and hence we may safely take $R_0 \approx b$ in \ref{ta21} to find dark energy's leading contribution. Also, using 
as $\phi \to 0$ in \ref{ta22}, we have  at the leading order,
$$\frac{1}{r}\approx \frac{2GM}{R_0^2}+\frac{15\pi M^2G^2}{8 R_0^3}$$
Taking $\tan\psi \approx \psi$ in \ref{ta19'} as earlier for small bending angle, we obtain the first order contribution due to the dark energy along with the usual terms,
$$
\psi \approx\frac{2GM}{R_0}+\frac{15\pi^2G^2M^2}{8R_0^2}+\frac{37G^3M^3}{8R_0^3}-\frac{\Lambda R_0^3}{12 GM}-\frac{6\Lambda b^5(1+\alpha)}{(GM)^3}
$$
And by virtue of the symmetry of the system about the line $\phi=\pi/2$, the total bending angle is just twice of the above 
\begin{eqnarray}
\Delta=2\psi \approx\frac{4GM}{R_0}+\frac{15\pi^2G^2M^2}{4R_0^2}+\frac{37G^3M^3}{4R_0^3}-\frac{\Lambda b^3}{6 GM}-\frac{12\Lambda b^5(1+\alpha)}{(GM)^3}
\label{ta23}
\end{eqnarray}
Thus for $\alpha\neq -1$ we have a new term in the expression for the bending angle and that term, as we have discussed, is completely due to $g_{tt}\cdot g_{rr}\neq -1$ in the metric we have dealt with. Note also that while the first of the dark energy terms is repulsive, the last term  becomes attractive for $\alpha<-1$.

Let us now estimate the relative strengths of the  leading terms in \ref{ta23} for the solar system, for which the first term equals $1.75$  arc seconds for $R_0\approx b\sim 10^7{\rm m}$~\cite{Wald:1984rg}. It is easy then to estimate that the ${\cal O}(GM/R_0)^2$ term is $\sim 10^{-4}$ times  the first. Whereas taking $\Lambda \sim 10^{-52}{\rm m}^{-2}$,  the first term due to the dark energy, which is identical to that of the Schwarzschild-de Sitter spacetime, is seen to be around $10^{-30}$ times the first. Coming to the last term, we find it is smaller than the first term by a factor of $\sim (1+\alpha)\times 10^{-22}$. Thus as expected, the dark energy has ignorable effect in the light bending in the solar system. However, we also note that modulo the $(1+\alpha)$ factor, the last term  is around $10^8$ times greater than the usual term due to the dark energy. Even though this does not count in the solar system anyway, one might expect to get interesting things at length scales of galaxies or clusters of galaxies, as we shall point out below.

\subsection{Light bending at large scales}\label{sec.4.1}
When one considers a `vacuole bounded by the static radius' description of a structure in $\Lambda{\rm CDM}$ ($\alpha=-1$) and uses the existing light bending data pertaining galaxies, while the leading contribution certainly happens to come from the first order mass term ($= 4MG/R_0$) in \ref{ta23}, the contribution due to the cosmological constant ($=-\Lambda R_0^3/6 GM$) turns out to be comparable with the second order contribution due to the mass ($=15\pi^2G^2M^2/4R_0^2 $), leading to an independent upper bound on $\Lambda$~\cite{Ishak:2007ea}.  Thus in order to be compatible with the observed data, undoubtedly the mass term  on the right hand side of \ref{ta23} must dominate the other terms. Now, since for $\alpha\neq -1$, the last term on the right hand side of \ref{ta23} is new, let us estimate it using certain large  scale light bending data.

But before we proceed, one caveat should be noted. In the case of the  cosmological strong lensing, one might need to consider light sources lying well outside the structure, i.e. the maximum turn around radius~\cite{Weinberg:2008zzc}. In that case clearly using the static description  and accordingly \ref{ta23} may be inappropriate. Note that this fact is also valid for the $\Lambda{\rm CDM}$, considered in~\cite{Ishak:2007ea}. In other words if we wish to fit the predictions of \ref{ta23} (irrespective of whether $\alpha =-1$), we must use sources located within the structure, such as a star in a galaxy, with the central overdensity playing the role of the lens. The fact that we had kept terms only linear in the dark energy in the expansion of the metric functions, \ref{add2}, also corresponds to this, i.e. we are interested in a length scale where the dark energy is not too dominant. Accordingly, we shall use the various data summarized and used in~\cite{Ishak:2007ea} (see also references therein) to constrain $\alpha$ appearing in \ref{ta23}.  Also, if we wish to apply \ref{ta23} for sources located outside the structure, we can make  a heuristic estimate on how far they could be to make our static approximation tentatively valid. Let us consider the spatially homogeneous part of \ref{nr2} where the proper or the physical radius is : $r=a(t)R$. Differentiating $r$ with respect to the time $t$ we have
$$v= H(t)r+a(t)\frac{dR}{dt}$$
where $v$ is the total relative speed between the light source and the lens, and $dR/dt$ is the radial peculiar speed. Thus we may apply the static approximation as long as $v\ll c$. Setting the current scale factor to unity and ignoring the peculiar speed, we have an upper bound, $r\lesssim c/H(t_0)$, where $H(t_0)$ is the current Hubble rate,   $\sim 73\, {\rm km\, s^{-1}Mpc^{-1}}$, yielding, $r_{\rm max}\lesssim {\cal O}(10^4) {\rm Mpc}$. Thus the proper distance between the source and the lens must lie {\it well within} this bound (say, for example, a couple of hundred Mpc), for \ref{ta23} to be safely applicable.  \\

\noindent All the data used below can be found in~\cite{Ishak:2007ea} and references therein.\\

For the {\it Abell 2744} galaxy cluster, we have $M \simeq 1.97 \times 10^{13}M_{\odot}h^{-1}$. Here $h$ is the dimensionless Hubble parameter defined as $H(t_0)=100h$, where $H(t_0)$ is the current Hubble rate $\sim 73\, {\rm km\, s^{-1}Mpc^{-1}}$, giving $h=0.73$. The leading deflection of the light for this system is measured to be $5.53\times 10^{-5}\,{\rm rad}$, whereas the subleading value is $2.25\times 10^{-9}\,{\rm rad}$. Equating the leading contribution to $4 MG/R_0$, we estimate $R_0 \simeq 1.01\times 10^{21}{\rm m}$. If we now plug these values into the last term on the right hand side of \ref{ta23}, using $\Lambda \sim 10^{-52}{\rm m^{-2}}$ and $R_0\approx b-GM \approx b$, we find it to be $\sim (1+\alpha)\times 10^{5}$! Clearly this is impossible to accept unless $\alpha =-1+\delta \alpha$, with $\delta \alpha$ is such that the contribution to the light bending due to this particular term is much small compared to the leading contribution and is comparable only with the second order value in the bending angle. This gives us a very strict bound : $|\delta \alpha| \lesssim   {\cal O}(10^{-14})$.  \\

For the {\it Abell 1689}  cluster,  $M\simeq 9.36 \times 10^{13}M_{\odot} h^{-1}$ and the leading value of the bending angle equals $1.88\times 10^{-4}\,{\rm rad}$ whereas the second order value equals $2.61\times 10^{-8}\,{\rm rad}$, giving $R_0= 1.4\times 10^{21}\,{\rm m}$. Plugging these values into the last term on the right hand side of \ref{ta23} as earlier, we find it equals $\sim (1+\alpha)\times 10^4$,   giving $ |\delta \alpha| \lesssim  {\cal O}(10^{-12})$.\\

For the {\it SDSS J1004 + 4112} quasar, we have $M\simeq 4.36\times 10^{13}M_{\odot}h^{-1}$ and the leading value of the bending angle is $1.05 \times 10^{-4}\,{\rm rad}$ whereas the second order value equals $8.06\times 10^{-9}\, {\rm rad}$, giving $R_0 \simeq 1.2\times 10^{21}\, {\rm m}$, yielding $|\delta \alpha| \lesssim {\cal O}(10^{-13})$. \\

For the {\it 3C 295 } cluster, we have $M \simeq 7.1\times 10^{13}M_{\odot }h^{-1}$, the leading and the next to the leading values of the bending angle equal $1.5\times 10^{-4}\,{\rm rad}$ and $1.66 \times 10^{-8}\,{\rm rad}$ respectively. This gives $R_0\simeq 1.4 \times 10^{21}\,{\rm m}$, yielding $|\delta \alpha| \lesssim {\cal O}(10^{-13})$.\\

For the {\it Abell 2219L} cluster, $M\simeq 3.22 \times 10^{13}M_{\odot }h^{-1}$, the leading and the next to the leading values of the bending angle equal $1.01\times 10^{-4}\,{\rm rad}$ and $7.47 \times 10^{-9}\,{\rm rad}$ respectively. This gives $R_0 \simeq 9.1\times 10^{20}\, {\rm m}$, yielding $|\delta \alpha| \lesssim  {\cal O}(10^{-14}) $.\\

Finally, for the {\it AC 114} cluster, $M\simeq 9.23 \times 10^{12}M_{\odot }h^{-1}$,  the leading and the next to the leading values of the bending angle equal $4.57\times 10^{-5}\,{\rm rad}$ and $1.54 \times 10^{-9}\,{\rm rad}$ respectively. This gives $R_0 \simeq 5.8\times 10^{20}\, {\rm m}$, yielding $|\delta \alpha| \lesssim  {\cal O}(10^{-13}) $.\\

\noindent
Putting these all in together, we conclude that in the scenario we are concerned with, we must have $(1+ \alpha) \lesssim {\cal O}(10^{-14})$, which is the main result of this paper. Certainly, such astonishingly small value was far from obvious {\it a priori}. We also note here that the inhomogeneity/staticity of the dark energy we have considered shares the same origin of the structure itself, e.g. \ref{ta3'}.  If on the other hand, the inhomogeneity is built up locally around a neighborhood of a point in a `small' region, and if the size of that region is much smaller than the impact parameter of the light ray, then certainly there would be ignorable
effect of such inhomogeneity into this phenomenon.   However, note that the description we are using is an effective one, to obtain a model of the static geometry in the entire bound zone of a structure. Given that the dark energy has repulsive effects, we have basically started  with the expectation that inhomogeneity/staticity of the dark energy, if any, should be built around a massive object/central overdensity owing to the equivalence principle. A different static metric (but with this same basic idea), as we have mentioned earlier, has recently been considered in~\cite{He:2017alg} in the context of light bending (see also references therein).

\section{Discussions}\label{sec.5}
\noindent
In this work we have considered a scenario to investigate and eventually to rule out, using the light bending phenomenon,  possible inhomogeneity of the dark energy modelled effectively in the form of an ideal static fluid, within the attraction dominated region of the maximum turn around radius of a cosmic structure.  We have also seen that as far as the non-relativistic motion of a test particle is concerned, it is analogous in \ref{ta0''} and the cosmological McVittie, \ref{nr2}. Next, we have used relevant data for the light bending  at the galactic scales to show that $\alpha$ is rather severely constrained to $-1$ :  $(1+\alpha)\lesssim {\cal O}(10^{-14})$.  Clearly such a tiny number was far from obvious {\it a priori}. While  certainly it is not expected  that the dark energy would build any considerable spatial inhomogeneity anyway, the above  estimate clearly probes  the severity of the `exclusion' of  it, as far as the effective fluid model we have considered is concerned.

 Now, as we have also discussed earlier,  the study of lensing in the McVittie background, \ref{nr2} or \ref{nr5}, in e.g.~\cite{Piattella:2015xga} (also references therein) yields  absolutely no effect  of the dark energy (including the $w=-1$ case) on the leading order bending angle. Thus we may fairly conclude that \ref{ta0''} and \ref{nr5} (or, \ref{nr2}) cannot be completely equivalent even though they predict equivalent trajectories in the non-relativistic sector (\ref{sec.2}), for in the relativistic sector the former predicts 
 that the dark energy is nothing but the cosmological constant in our effective static description, $(1+\alpha)\lesssim {\cal O}(10^{-14})$, with a non-vanishing effect on the light bending angle, \ref{ta23}. Hence perhaps it would be an interesting task to compute the post Newtonian corrections~\cite{Will} of \ref{nr1} and \ref{nr8} and to see from what precise magnitude of the velocity of the test particle, we start obtaining considerable disagreement between the predicted trajectories. Perhaps our analysis thus adds to the recent debate on whether or under what conditions a McVittie geometry can be equivalent to a static description, inside the maximum turn around radius~\cite{Skordis:2017qxp} (also references therein).

Finally, we have shown that,  inside $R_{\rm TA, max}$,   as far as obtaining a static spacetime geometry is concerned with an effective static and ideal dark energy fluid, the Schwarzschild-de Sitter spacetime is the only acceptable solution. Since this result is basically established by ruling out any such  dark energy fluid other than a positive $\Lambda$, we believe it to be interesting in its own right.  Note that this is not analogous to the proof of Birkhoff's theorem with $\Lambda>0$~\cite{Schleich}, as we have essentially not dealt with isolated stationary de Sitter black holes here.

\bigskip
\section*{Acknowledgements}
\noindent
We acknowledge S.~Chakraborty, N.~K.~Dadhich, A.~Lahiri and K.~Lochan for  discussions. We would also like to thank S.~G.~Ghosh for his comments on an earlier version of the manuscript. We also sincerely thank
anonymous referees for various useful comments.



\bigskip


\begin{thebibliography}{99} 


\bibitem{Bhattacharya:2017yix} 
  S.~Bhattacharya and T.~N.~Tomaras,
  ``{\it Cosmic structure sizes in generic dark energy models}'',
  Eur.\ Phys.\ J.\ C {\bf 77}, no. 8, 526 (2017)
  arXiv:1703.07649 [gr-qc].
  
   
\bibitem{Wald:1984rg}
R.~M.~Wald, ``{\it General Relativity}'', Chicago Univ. Press (1986)


  
  \bibitem{Paddy}
   T.~Padmanabhan, ``{\it Gravitation : Foundation and Frontiers}'', Cambridge Univ. Press (2010)
   
   
     
\bibitem{Will}
C.~M.~Will, ``{\it Theory and Experiment in Gravitational Physics}'', Cambridge Univ. Press (1993).
   

   
\bibitem{Bjerrum-Bohr:2014zsa} 
  N.~E.~J.~Bjerrum-Bohr, J.~F.~Donoghue, B.~R.~Holstein, L.~Planté and P.~Vanhove,
  ``{\it Bending of Light in Quantum Gravity}'',
  Phys.\ Rev.\ Lett.\  {\bf 114}, no. 6, 061301 (2015)
  [arXiv:1410.7590 [hep-th]].   
   
   
   
\bibitem{Rodrigues:2015rya} 
  D.~C.~Rodrigues, B.~Koch, O.~F.~Piattella and I.~L.~Shapiro,
  ``{\it The bending of light within gravity with large scale renormalization group effects}'',
  AIP Conf.\ Proc.\  {\bf 1647}, 57 (2015).   
   
 
\bibitem{Weinberg:2008zzc} 
  S.~Weinberg,
  ``{\it Cosmology}'', Oxford Univ. Press (2008)

 
 
 
 \bibitem{Finelli:2006iz}
  F.~Finelli, M.~Galaverni and A.~Gruppuso,
  ``{\it Light bending as a probe of the nature of dark energy}'',
  Phys.\ Rev.\ D {\bf 75}, 043003 (2007)
  [astro-ph/0601044].
 
  
  \bibitem{Rindler:2007zz}  
  W.~Rindler and M.~Ishak,
  ``{\it Contribution of the cosmological constant to the relativistic bending of light revisited}'',
  Phys.\ Rev.\ D {\bf 76}, 043006 (2007)
  [arXiv:0709.2948 [astro-ph]].
  
  
  \bibitem{Ishak:2007ea} 
  M.~Ishak, W.~Rindler, J.~Dossett, J.~Moldenhauer and C.~Allison,
  ``{\it A New Independent Limit on the Cosmological Constant/Dark Energy from the Relativistic Bending of Light by Galaxies and Clusters of Galaxies}'',
  Mon.\ Not.\ Roy.\ Astron.\ Soc.\  {\bf 388}, 1279 (2008)
  [arXiv:0710.4726 [astro-ph]].

 
 
 \bibitem{Sereno:2007rm} 
  M.~Sereno,
  ``{\it On the influence of the cosmological constant on gravitational lensing in small systems}'',
  Phys.\ Rev.\ D {\bf 77}, 043004 (2008)
  [arXiv:0711.1802 [astro-ph]].
 
 
 
 \bibitem{Schucker:2007ut} 
  T.~Schucker,
  ``{\it Cosmological constant and lensing}'',
  Gen.\ Rel.\ Grav.\  {\bf 41}, 67 (2009)
  [arXiv:0712.1559 [astro-ph]].
 
 
  
  
  \bibitem{Ishak:2008ex} 
  M.~Ishak,
  ``{\it Light Deflection, Lensing, and Time Delays from Gravitational Potentials and Fermat's Principle in the Presence of a Cosmological Constant}'',
  Phys.\ Rev.\ D {\bf 78}, 103006 (2008)
  [arXiv:0801.3514 [astro-ph]].
  
  
  \bibitem{Schucker:2008bc} 
  T.~Schucker,
  ``{\it Strong lensing in the Einstein-Straus solution}'',
  Gen.\ Rel.\ Grav.\  {\bf 41}, 1595 (2009)
  [arXiv:0807.0380 [astro-ph]].
  
  
  \bibitem{Sereno:2008kk} 
  M.~Sereno,
  ``{\it The role of Lambda in the cosmological lens equation}'',
  Phys.\ Rev.\ Lett.\  {\bf 102}, 021301 (2009)
  [arXiv:0807.5123 [astro-ph]].
  
   
  \bibitem{Schucker:2009ke} 
  T.~Schucker,
  ``{\it Lensing in an interior Kottler solution}'',
  Gen.\ Rel.\ Grav.\  {\bf 42}, 1991 (2010)
  [arXiv:0903.2940 [astro-ph.CO]].
  
  \bibitem{Arakida:2011ty} 
  H.~Arakida and M.~Kasai,
  ``{\it Effect of the cosmological constant on the bending of light and the cosmological lens equation}'',
  Phys.\ Rev.\ D {\bf 85}, 023006 (2012)
  [arXiv:1110.6735 [gr-qc]].
  
  
  \bibitem{Bhadra:2010jr} 
  A.~Bhadra, S.~Biswas and K.~Sarkar,
  ``{\it Gravitational deflection of light in the Schwarzschild -de Sitter space time}'',
  Phys.\ Rev.\ D {\bf 82}, 063003 (2010)
  [arXiv:1007.3715 [gr-qc]].
  
  
  \bibitem{Kantowski:2009jt} 
  R.~Kantowski, B.~Chen and X.~Dai,
  ``{\it Gravitational Lensing Corrections in Flat $\Lambda$CDM Cosmology}''
  Astrophys.\ J.\  {\bf 718}, 913 (2010)
  [arXiv:0909.3308 [astro-ph.CO]].
  
  
  
  
  \bibitem{Lebedev:2016kun} 
  D.~Lebedev and K.~Lake,
  ``{\it Relativistic Aberration and the Cosmological Constant in Gravitational Lensing I: Introduction}'',
  arXiv:1609.05183 [gr-qc].
  
  
  
    
 \bibitem{Bambi:2007hj} 
  C.~Bambi,
  ``{\it Dark Energy and the mass of galaxy clusters}'',
  Phys.\ Rev.\ D {\bf 75}, 083003 (2007)
  [astro-ph/0703645]. 
  
  
  
  
  \bibitem{Bhattacharya:2010xh} 
  A.~Bhattacharya, G.~M.~Garipova, E.~Laserra, A.~Bhadra and K.~K.~Nandi,
  ``{\it The Vacuole Model: New Terms in the Second Order Deflection of Light}'',
  JCAP {\bf 1102}, 028 (2011)
  [arXiv:1002.2601 [gr-qc]].
   
  
  
  
  
  
  
  \bibitem{Kraniotis:2010gx}
  G.~V.~Kraniotis,
  ``{\it Precise analytic treatment of Kerr and Kerr-(anti) de Sitter black holes as gravitational lenses}'',
  Class.\ Quant.\ Grav.\  {\bf 28}, 085021 (2011)
  [arXiv:1009.5189 [gr-qc]].
  
  \bibitem{Sultana:2013ppa} 
  J.~Sultana,
  ``{\it Contribution of the cosmological constant to the bending of light in Kerr–de Sitter spacetime}'',
  Phys.\ Rev.\ D {\bf 88}, no. 4, 042003 (2013).
  
  
  \bibitem{Fernando:2014rsa} 
  S.~Fernando, S.~Meadows and K.~Reis,
  ``{\it Null trajectories and bending of light in charged black holes with quintessence}'',
  Int.\ J.\ Theor.\ Phys.\  {\bf 54}, no. 10, 3634 (2015)
  [arXiv:1411.3192 [gr-qc]].
  
  
  \bibitem{Soroushfar:2016yea} 
  S.~Soroushfar, R.~Saffari and E.~Sahami,
  ``{\it Geodesic equations in the static and rotating dilaton black holes: Analytical solutions and applications}'',
  Phys.\ Rev.\ D {\bf 94}, no. 2, 024010 (2016)
  [arXiv:1601.03143 [gr-qc]].
  
  
  \bibitem{Iorio:2016sqy} 
  L.~Iorio, M.~L.~Ruggiero, N.~Radicella and E.~N.~Saridakis,
  ``{\it Constraining the Schwarzschild–de Sitter solution in models of modified gravity}'',
  Phys.\ Dark Univ.\  {\bf 13}, 111 (2016)
  [arXiv:1603.02052 [gr-qc]].
  

  \bibitem{Soroushfar:2016esy} 
  S.~Soroushfar, R.~Saffari, S.~Kazempour, S.~Grunau and J.~Kunz,
  ``{\it Detailed study of geodesics in the Kerr-Newman-(A)dS spacetime and the rotating charged black hole spacetime in $f(R)$ gravity}'',
  Phys.\ Rev.\ D {\bf 94}, no. 2, 024052 (2016)
  [arXiv:1605.08976 [gr-qc]].
  
  
  
  \bibitem{Zhao:2016ltm} 
  F.~Zhao, J.~Tang and F.~He,
  ``{\it Gravitational lensing effects of a Reissner–Nordstrom–de Sitter black hole}'',
  Phys.\ Rev.\ D {\bf 93}, no. 12, 123017 (2016).
 
 
 
 
  
   
  \bibitem{Villanueva:2013gga}
  J.~R.~Villanueva and M.~Olivares,
  ``{\it On the Null Trajectories in Conformal Weyl Gravity}'',
  JCAP {\bf 1306}, 040 (2013)
  [arXiv:1305.3922 [gr-qc]].
  
  \bibitem{Olivares:2016zqq} 
  M.~Olivares, C.~Osses and J.~R.~Villanueva,
  ``{\it Photon paths around hyperbolic topological black holes in conformal Weyl gravity}'',
  arXiv:1608.02979 [gr-qc].
  
  
  \bibitem{Lim:2016lqv} 
  Y.~K.~Lim and Q.~h.~Wang,
  ``{\it Exact gravitational lensing in conformal gravity and Schwarzschild-de Sitter spacetime}'',
  Phys.\ Rev.\ D {\bf 95}, no. 2, 024004 (2017)
  [arXiv:1609.07633 [gr-qc]].
  
    
  
  \bibitem{Virbhadra:2007kw} 
  K.~S.~Virbhadra and C.~R.~Keeton,
  ``{\it Time delay and magnification centroid due to gravitational lensing by black holes and naked singularities}''
  Phys.\ Rev.\ D {\bf 77}, 124014 (2008)
  [arXiv:0710.2333 [gr-qc]].
 
  
  
  
   \bibitem{Bhattacharya:2008fu} 
  S.~Bhattacharya and A.~Lahiri,
  ``{\it Effect of a positive cosmological constant on cosmic strings}'',
  Phys.\ Rev.\ D {\bf 78}, 065028 (2008)
  [arXiv:0807.0543 [gr-qc]]
  
  
  \bibitem{Khriplovich:2008ij} 
  I.~B.~Khriplovich and A.~A.~Pomeransky,
  ``{\it Does Cosmological Term Influence Gravitational Lensing?}'',
  Int.\ J.\ Mod.\ Phys.\ D {\bf 17}, 2255 (2008)
  [arXiv:0801.1764 [gr-qc]].
   
  \bibitem{Simpson:2008jf} 
  F.~Simpson, J.~A.~Peacock and A.~F.~Heavens,
  ``{\it On lensing by a cosmological constant}'',
  Mon.\ Not.\ Roy.\ Astron.\ Soc.\  {\bf 402}, 2009 (2010)
  [arXiv:0809.1819 [astro-ph]].
  
  
  \bibitem{Butcher:2016yrs} 
  L.~M.~Butcher,
  ``{\it No practical lensing by Lambda: Deflection of light in the Schwarzschild–de Sitter spacetime}'',
  Phys.\ Rev.\ D {\bf 94}, no. 8, 083011 (2016)
  [arXiv:1602.02751 [gr-qc]].
  
 
  
 \bibitem{Piattella:2015xga} 
  O.~F.~Piattella,
  ``{\it Lensing in the McVittie metric}'',
  Phys.\ Rev.\ D {\bf 93}, no. 2, 024020 (2016)
  Erratum: [Phys.\ Rev.\ D {\bf 93}, no. 12, 129901 (2016)]
  [arXiv:1508.04763 [astro-ph.CO]].
  
  \bibitem{Faraoni:2016wae} 
  V.~Faraoni and M.~Lapierre-Leonard,
  ``{\it Beyond lensing by the cosmological constant}'',
  Phys.\ Rev.\ D {\bf 95}, no. 2, 023509 (2017)
  [arXiv:1608.03164 [gr-qc]].
 
 
 \bibitem{He:2017alg} 
  H.~J.~He and Z.~Zhang,
  ``{\it Direct Probe of Dark Energy through Gravitational Lensing Effect}'',
  JCAP {\bf 1708}, no. 08, 036 (2017)
  arXiv:1701.03418 [astro-ph.CO].
  
  
  
   \bibitem{Chamseddine:2016uyr} 
  A.~H.~Chamseddine and V.~Mukhanov,
  ``{\it Inhomogeneous Dark Energy}'',
  JCAP {\bf 1602}, no. 02, 040 (2016)
  [arXiv:1601.04941 [astro-ph.CO]].
  
  
  \bibitem{Kluson:2017iem} 
  J.~Kluson,
  ``{\it Canonical Analysis of Inhomogeneous Dark Energy Model and Theory of Limiting Curvature}'',
  JHEP {\bf 1703}, 031 (2017)
  [arXiv:1701.08523 [hep-th]].
  
   \bibitem{Sebastiani:2016ras} 
  L.~Sebastiani, S.~Vagnozzi and R.~Myrzakulov,
  ``{\it Mimetic gravity: a review of recent developments and applications to cosmology and astrophysics}'',
  Adv.\ High Energy Phys.\  {\bf 2017}, 3156915 (2017)
  [arXiv:1612.08661 [gr-qc]].

  
 
  
  \bibitem{Stuclik1}
Z.~Stuchlik,
``{\it The motion of test particles in black-hole backgrounds with non-zero cosmological constant}'',
Bull.~Astronom. Inst. Czechoslovakia {\bf 34}, no. 3,  129 (1983)

\bibitem{Stuclik2}
Z.~Stuchlik, P.~Slany and S.~Hledik,
``{\it Equilibrium configurations of perfect fluid orbiting Schwarzschild-de Sitter black holes}'',
Astronomy \& Astrophysics{\bf 363},  no. 2,  425 (2000)



\bibitem{Pavlidou:2013zha} 
  V.~Pavlidou and T.~N.~Tomaras,
  ``{\it Where the world stands still: turnaround as a strong test of $\Lambda$CDM cosmology}'',
  JCAP {\bf 1409}, 020 (2014)
  [arXiv:1310.1920 [astro-ph.CO]]



\bibitem{Pavlidou:2014aia} 
  V.~Pavlidou, N.~Tetradis and T.~N.~Tomaras,
  ``{\it Constraining Dark Energy through the Stability of Cosmic Structures}'',
  JCAP {\bf 1405}, 017 (2014)
  [arXiv:1401.3742 [astro-ph.CO]]
  
  
  \bibitem{Hannestad}
S.~Hannestad, ``{\it Constraints on the sound speed of dark energy}",
Phys.\ Rev. \ D {\bf 71}, 103519 (2005)
[astro-ph/0504017]
  

\bibitem{Jacobson:2007tj} 
  T.~Jacobson,
  ``{\it When is $g_{tt}\, g_{rr} = -1$?}'',
  Class.\ Quant.\ Grav.\  {\bf 24}, 5717 (2007)
  [arXiv:0707.3222 [gr-qc]].
  
  \bibitem{Skordis:2017qxp} 
  C.~Skordis,
  ``{\it Conditions for equivalence of static spherically symmetric spacetimes to almost Friedman-Robertson-Walker}'',
  arXiv:1708.08797 [gr-qc].
  
  \bibitem{Schleich}
  K.~Schleich and D.~M.~Witt, ``{\it A simple proof of Birkhoff's theorem for cosmological constant}", J.\ Math. \ Phys.\ {\bf 51}, 112502 (2010) [arXiv:0908.4110 [gr-qc]]
  

  



\end{thebibliography}
\end{document}